\documentclass[aps,pre,twocolumn,showpacs]{revtex4}

\usepackage[dvips]{graphicx}
\usepackage{amssymb}
\usepackage{amsmath}

\begin{document}

\title{Drainage of a nanoconfined simple fluid: rate effects on squeeze-out dynamics}

\author{Lionel Bureau}
\email{bureau@insp.jussieu.fr}
\author{Arnaud Arvengas}
\affiliation{CNRS, UMR7588, INSP, Paris, Campus Boucicaut, 140 rue de Lourmel, Paris, F-75015 France}
\affiliation{Universit\'e Pierre et Marie Curie-Paris6, UMR 7588, INSP, Paris, F-75015 France}

\date{\today}

\begin{abstract}

We investigate the effect of loading rate on drainage in molecularly thin films of a simple fluid made of quasi-spherical molecules (octamethylcyclotetrasiloxane, OMCTS). 
We find that (i) rapidly confined OMCTS retains its tendency to organize into layers parallel to the confining surfaces, and (ii) flow resistance in such layered films can be 
described by bulklike viscous forces if one accounts for the existence of one monolayer immobilized on each surfaces. The latter result is fully consistent with the recent work 
of Becker and Mugele, who  reached a similar conclusion by analyzing the dynamics of squeeze-out fronts in OMCTS [T. Becker and F. Mugele, Phys. Rev. Lett. {\bf 91} 
166104(2003)]. Furthermore, we show that the confinement rate controls the nature of the thinning transitions: layer-by-layer expulsion of molecules in metastable, slowly 
confined films proceeds by a nucleation/growth mechanism, whereas deeply and rapidly quenched films are unstable and undergo thinning transitions akin to spinodal 
decomposition.

\end{abstract}


\maketitle

\section{Introduction}
\label{sec:intro}

Ultrathin liquid films are ubiquitous in situations involving wetting and elaboration of coatings on solid substrates \cite{wet1,wet2,wet3}, manipulation of fluids into 
nanometer-sized channels (nanofluidics) \cite{nano1,nano2,CB}, 
or boundary lubrication of frictional contacts, where two solid surfaces in relative motion are separated by a few monolayers of lubricant \cite{bo,IMH1,TR}. 
Stability and flow properties of such molecularly thin films therefore represent key issues in these various fields. 

The surface forces apparatus (SFA) allows for studies of ultrathin films under extremely well characterized conditions of thickness and applied pressure \cite{IA,TW,IT}. It is
therefore a choice tool for experimental investigations of the structural and mechanical properties of strongly confined liquids. 
SFA studies have thus provided evidence that in the 
vicinity of a solid wall, a fluid tends to order and form molecular layers parallel to the surface, such a wall-induced structure 
extending a few nanometers away from the interface 
into the fluid \cite{Ibook}. When two atomically smooth solids are approached down to nanometer separations,  the layered structure of the intercalated liquid 
 gives rise to the so-called solvation forces: the interaction 
force between the surfaces oscillates with the distance, and alternates between repulsive maxima and attractive minima with a periodicity comparable to the molecular 
diameter of the confined liquid \cite{HI,HI2,HKC1}. 

Such a layering effect has a strong impact on the flow properties of nm-thick films. It has been shown for instance that a structured liquid exhibits an effective viscosity which is markedly enhanced with respect to the 
bulk one \cite{DG,RDG,KK1,KK2,JPA}. Most of these studies were
concerned with the properties of ultrathin films formed under quasi-static compression, so as to probe the behavior of systems having reached their equilibrium layered structure. \\
However, boundary lubrication commonly involves the approach at finite velocity of the surfaces, which implies that molecularly-thin lubricant films are {\it dynamically} loaded.
 An important issue is therefore to determine how layering, and the accompanying changes in flow properties, 
may be affected by the velocity at which the liquid is ``mechanically quenched'' \cite{ZG,ZG2}.

We recently addressed this question by performing, on a home-built SFA, drainage experiments at various velocities, using n-hexadecane (a linear alkane) as the confined fluid \cite{LB1}. We have shown that under rapid quenching 
conditions, the repulsive maxima characterizing solvation forces disappear, indicating that  layering was  hindered or disrupted. Moreover, our results suggested that such disordered films of hexadecane, exhibiting 
enhanced
viscosity, were  metastable and relaxed towards their layered configuration, most likely via a nucleation/growth process.

Several questions arise from this study : (i) to what extent does the molecular architecture of the confined lubricant influence its behavior under rapid quenching ? (ii) which mechanisms govern the squeeze-out and 
thinning of the lubricant, and how are they affected by the confinement rate ? 

These questions have led us to  extend our previous study to the case of a fluid of simpler molecular geometry, namely octamethylcyclotetrasiloxane (OMCTS, a quasi-spherical molecule).
We find that, in contradistinction with n-hexadecane, (i) rapidly confined OMCTS still exhibits layering, and (ii) the magnitude of the repulsive forces measured during drainage can be accounted for by the 
combination
of solvation and hydrodynamic forces. Analysis of the data obtained under conditions where viscous drag dominates over surface forces suggests that {\it layered} OMCTS retains its bulk viscosity if one accounts, 
as proposed in earlier studies \cite{CH}, 
for the existence of one monolayer immobilized on each surface.  
This result is fully consistent with the recent work of Becker and Mugele, who  reached a similar conclusion by analyzing the dynamics of squeeze-out fronts in OMCTS \cite{BM}.

In a spirit similar to the work of these authors, we have performed squeeze-flow experiments while imaging the confined film. This allows us to investigate the effect of loading conditions on the nature of thinning 
transitions. \\
We observe that, when
the pressure applied on a layered film of OMCTS is  slowly ramped up, single monolayer squeeze-out occurs by a nucleation/growth mechanism, as already evidenced by 
Mugele {\it et al} \cite{BM,BM2,MS}. However, 
 we observe that
 two different scenarii exist: nucleation at the center of the film, followed by outward propagation of a front, or nucleation at the periphery of
 the confined film and subsequent propagation of an inward front.
 Analysis of
 these two modes of propagation allows us to (i) give further support to a model proposed by Persson and Tosatti \cite{PT}, and (ii) precise the role of line tension
 in nucleation of a thin zone inside a confined film. \\
Moreover, we show that when the pressure applied to the lubricant is increased {\it abruptly}, thinning from $n$ to $n-1$ layers occurs by a process akin to spinodal decomposition: 
multiple zones of thickness $n-1$ appear simultaneously, grow and coalesce to finally leave a film of homogeneous thickness $n-1$.

These results indicate that the rate of increase and the magnitude of the pressure applied on a confined film of OMCTS determine the dynamics of layering transitions. Squeeze-out can occur either by nucleation/growth or by 
layer-by-layer ``spinodal thinning'', 
depending on the depth of the mechanical quench, in strong analogy with thermal quenching of binary mixtures \cite{beysens,tanaka} or dewetting of ultrathin polymer films \cite{dewet}.

\section{Experiments}
\label{sec:exp}

The experiments were performed using a home-built surface force apparatus described elsewhere \cite{LB}. 

Two atomically smooth mica sheets, mounted in a crossed-cylinder
 geometry, are used to confine the liquid under study.   The thickness $d$ of the liquid is determined using multiple beam interferometry : the backside of the mica sheets is made highly reflective by evaporation
 of a thin silver film, and the mica/liquid/mica layered medium  forms a Fabry-Perot interferometer which is shone with a collimated beam of white light. The Fabry-Perot cavity transmits a set of discrete wavelengths,
 corresponding to the so-called fringes of equal chromatic order (FECO). Analysis of the spectral position of the FECO allows to determine the thickness of the fluid between the mica surfaces .
 The normal force $F$ applied on the confined fluid is measured by means of a capacitive load cell of stiffness 9500 N.m$^{-1}$. 
Force and thickness can thus be measured independently at a rate of up to 30 Hz. \\
 This allows us to obtain $F(d)$ curves during drainage of the
 fluid, while the mica surfaces are approached by driving the remote point of the loading spring at a prescribed velocity $V$ in the range 5.10$^{-2}$--10$^4$ nm.s$^{-1}$.
 
The mica sheets were prepared according to the following protocole. Muscovite mica plates (JBG-Metafix, France) were cleaved down to a thickness of $\sim$ 10 $\mu$m, cut
into 1 cm$^{2}$ samples by means of surgical scissors, and coated on one side with a 40nm-thick thermally evaporated silver layer. The sheets were fixed, silver side down, onto cylindrical glass lenses
(radius of curvature $R\simeq 1$ cm), using a
soft UV setting glue (NOA 81, Norland). The mica sheets were 
placed on the lenses so that their crystallographic axis be aligned. Prior to the experiments, each mica sample was recleaved using adhesive tape \cite{FS}, and mounted in the apparatus
so that the region of closest distance between the lenses be free of steps. The surfaces were brought into contact under an argon atmosphere, and the total mica thickness was
deduced from the position of the FECO using the multilayer matrix method \cite{Heuberg1,LB}. The surfaces were then separated and the thickness of each mica sheet
determined using the same method. A drop of liquid ($\sim$30--50 $\mu$L) was finally injected between the surfaces, a beaker containing P$_{2}$O$_{5}$ was placed inside the apparatus which was then sealed
and left for thermal equilibration for about 6 hours before measurements began.

The liquid used was octamethylcyclotetrasiloxane (OMCTS, Fluka purum $>$99\%). OMCTS is a slightly oblate molecule of major diameter $\sim$ 1 nm and minor diameter $\sim 0.8$ nm. 
 The product was used as received and filtered through a 0.2 $\mu$m teflon membrane immediately before injection.
All the experiments reported below have been performed at  $T=22 \pm 0.02^{\circ}$C.

\section{Results and discussion}
\label{sec:results}

\subsection{Squeeze flow of confined OMCTS : force-distance curves}
\label{subsec:drain}

In a first series of experiments, we have measured force-distance curves during approach of the mica surfaces at different driving velocities, starting from separations on the order of 100 nm. This was done using thick mica sheets (thickness $\sim 5 \mu$m). 
Under these conditions, the effective stiffness of the  mica sheet/glue layer composite is such that no dicernible change in the radius of curvature of the mica sheets is detected for normal loads below
$10^{-4}$ N. Surface flattening occurs under larger forces. 

\begin{figure}[htbp]
$$
\includegraphics[width=8cm]{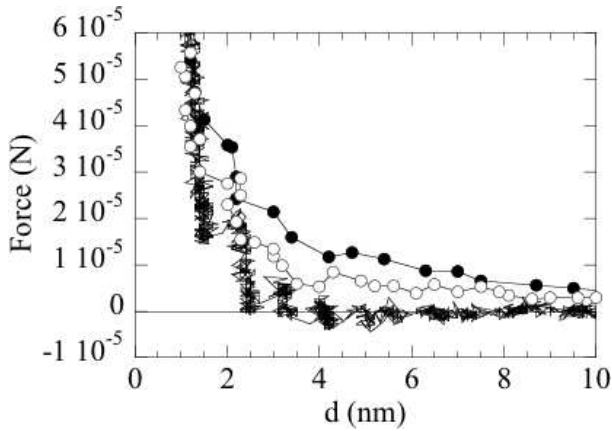}
$$
\caption{Force $vs$ distance curves measured during approach at $V=0.1$ nm.s$^{-1}$ (full line); $V=5$ nm.s$^{-1}$ (full line with $\circ$ markers);
$V=10$ nm.s$^{-1}$ (full line with $\bullet$ markers).
}
\label{fig:fig1}
\end{figure}

$F(d)$ profiles measured at $V=0.1$, 5 and 10 nm$.s^{-1}$ are plotted on Fig. \ref{fig:fig1}. Layering of OMCTS is clearly visible on the approach curve obtained at the slowest driving velocity : for intersurface distances
below  $\sim$6 nm, the thickness of the liquid film decreases by steps of approximately 1 nm while the applied load increases. Approaching at higher driving velocities results in an upward shift of the $F(d)$ curves, all the more so 
that $V$ is larger. \\
Although less well resolved because of acquisition rate limitations, step-like thinning can still be distinguished at $V=5$ nm$.s^{-1}$ (for $d<4$ nm) and $V=10$ nm$.s^{-1}$ ($d<3$ nm). These results suggest that, even under 
dynamic loading,  confined OMCTS still exhibits some layering. We will show later that layering is indeed preserved even at very large driving velocities (V$\gtrsim$ 500 nm.s$^{-1}$). This is in marked contrast with what we have observed for the linear 
alkane n-hexadecane, which, under similar loading conditions, behaves as a disordered fluid for driving velocities above 
$\sim$1 nm.$s^{-1}$.

\begin{figure}[htbp]
$$
\includegraphics[width=8cm]{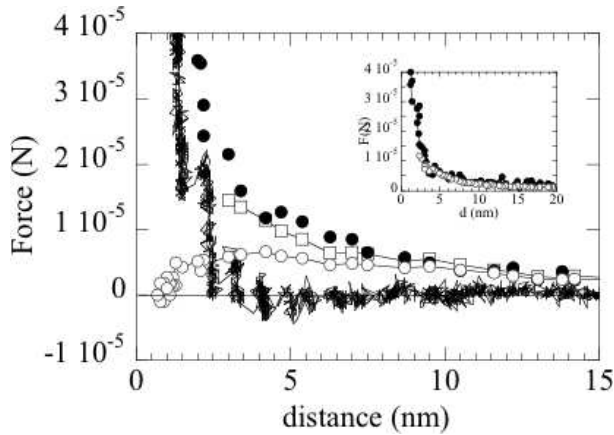}
$$
\caption{Main panel : force $vs$ distance curves measured during approach at $V=0.1$ nm.s$^{-1}$ (full line) and $V=10$ nm.s$^{-1}$ ( $\bullet$ symbols). Reynolds force computed for
a no-slip plane located at the surfaces (line with $\circ$ symbols), and a no-slip plane shifted by two monolayers ($h=1.8$ nm) (line with $\square$ symbols).
Inset: $F(d)$ measured at $V=5$ nm.s$^{-1}$ ($\bullet$) and $F_{H}$ computed with $h=1.8$ nm (line with $\circ$ symbols).
}
\label{fig:fig2}
\end{figure}

Moreover, the shift of the $F(d)$ curves towards higher forces at larger velocities can be accounted for by hydrodynamic forces, {\it i.e.} by viscous drag associated with the Poiseuille flow induced when the surfaces 
are brought together. From the measurement of $d$ as a function of time, we compute 
$\dot{d}$, the true approach velocity of the surfaces (which departs from $V$ as $F$ increases), and thus evaluate the Reynolds force:
\begin{equation}
\label{eq:reynolds}
 F_{H}=6\pi \eta R^{2}\frac{\dot{d}}{(d-h)}
 \end{equation}
 where $\eta$  is the viscosity of OMCTS, $R=10^{-2}$ m is the radius of curvature of the mica sheets, and the distance $h$ can be chosen in order to shift the no-slip boundary 
 condition.\\
 Comparison of the $F(d)$ curves measured at $V=5$ and 10 nm.s$^{-1}$ with $F_{H}$ is presented on Fig. \ref{fig:fig2}. 
 $F_{H}$ calculated using the {\it bulk viscosity}, $\eta\simeq 2.10^{-3}$ Pa.s, and a no-slip boundary condition at the mica surface ($h=0$) accounts well for the measured repulsive force for distances larger 
 than $\sim$10 nm, but departs noticeably from  the measured profile for thinner films. Chan and Horn, in a pioneer study, noticed that drainage in confined films of thickness down $\sim$5--6 nm was satisfactorily described 
 by simply assuming that the monolayers in direct contact with the surfaces are ``immobile'' \cite{CH}. This amounts to shift the no-slip boundary condition inside the gap, and use $h=2a$
 in eq. \ref{eq:reynolds} ($a$ being the molecular diameter).  \\
 Following the conclusions of Chan and Horn, we have computed $F_{H}$ as a function of $d$ using $h=1.8$ nm, while keeping for $\eta$ the value of the bulk viscosity. It can be seen
 on Fig. \ref{fig:fig2} that the hydrodynamic force calculated under these assumptions  accounts quite well for the average upward shift of the $F(d)$ curves measured at large driving velocities,
  down to a thickness on the order of 3 monolayers.\\
  Such an analysis therefore indicates that viscous drag in {\it layered films} of OMCTS ($d\lesssim 6$ nm) is well described by the {\it bulk viscosity} of the fluid, provided that one accounts for one immobile
  monolayer on each confining wall.

 This set of results leads us to the following conclusions:
 
 (i) the confinement rate does not affect significantly the tendency of OMCTS to form layers close to the walls. This indicates that the characteristic time scale of the 
 drainage flows (the inverse of the shear rate) remains larger than the relaxation time of the confined fluid,
 
 (ii) the measured repulsive forces result from the combination of solvation and hydrodynamic forces, their respective weight depending on the confinement rate,
 
 (iii) The analysis of our approach curves obtained at large velocities, where hydrodynamic forces are significant, is fully consistent with the results of Becker and Mugele (BM) in
 their experiments on squeeze-out fronts dynamics \cite{BM}. These authors indeed reached the conclusion that
 the flow resistance of confined OMCTS can be attributed to bulk-like viscous friction between adjacent layers in relative motion, combined to much higher dissipation arising from friction of the monolayers in direct
 contact with the confining walls.
 
 (iv)  The flow properties of confined OMCTS contrast noticeably with those of n-hexadecane. Indeed, we have found, in a previous study, that rapidly confined hexadecane exhibited a flow resistance that could not be 
 accounted for by bulk-like hydrodynamic forces \cite{LB1}. Analysis of the drainage curves suggested that when rapidly quenched, the linear alkane behaved as a disordered fluid, which viscosity increased by more than one order of
 magnitude as its thickness was reduced from 10 to 2 nm. We believe that these marked differences in the behaviors of fluids made of quasi-spherical or linear molecules arise from the possibility, for long enough 
 chain molecules, to form entanglements, leading, under confinement, to dynamical slowdown. This underlines the 
 importance of molecular architecture in the flow of dynamically confined liquids.
 
\subsection{Slowly confined films: thinning by nucleation/growth}
\label{subsec:fronts}

We have further investigated how the loading rate may affect the dynamics of layering transitions. For this purpose, we have performed confinement experiments using thin mica sheets, whose thickness was
on the order of a few hundreds of nanometers \cite{footnote} (namely between 300 and 800 nm. Such thin sheets are obtained by successive cleavage using adhesive tape). Under such conditions, the mica/glue compound readily flattens under low applied loads
($F\lesssim$5.10$^{-5}$ N), and forms a contact of finite area that can reach hundreds of micrometers in diameter. \\
In parallel to the thickness measurement of the liquid confined into this slit pore, which is performed by white light multiple beam interferometry, we image the contact area by directing half of the transmitted 
light intensity towards a 2D CCD camera. Doing so, the Fabry-Perot interference patterns associated with the different transmitted wavelengths overlap on the CCD chip, and the resulting intensity does not
directly reflect the thickness of the confined medium, unlike in the monochromatic SFA experiments of Mugele {\it et al} \cite{MS,BM2}. 

When imaging a contact containing a liquid 
film of uniform thickness, we observe a uniform transmitted intensity, except at the periphery of the flattened area. 
Starting from a homogeneous film having a thickness of $n$ monolayers, we ramp up the applied pressure by continuously driving the remote point of the loading cell at constant velocity, while
imaging the contact area.

Depending on the amplitude and rate of increase of the applied load, we observed three different types of scenarii for thinning transitions. 

\subsubsection{Outward fronts}
\label{subsubsec:out}

In experiments where the applied load is ramped up using moderate driving velocities ($V=5$--20 nm.$s^{-1}$, which translates into loading rates of $3.10^{-5}$--$12.10^{-5}$ N.s$^{-1}$), we observe the following.

\begin{figure}[htbp]
$$
\includegraphics[width=8cm]{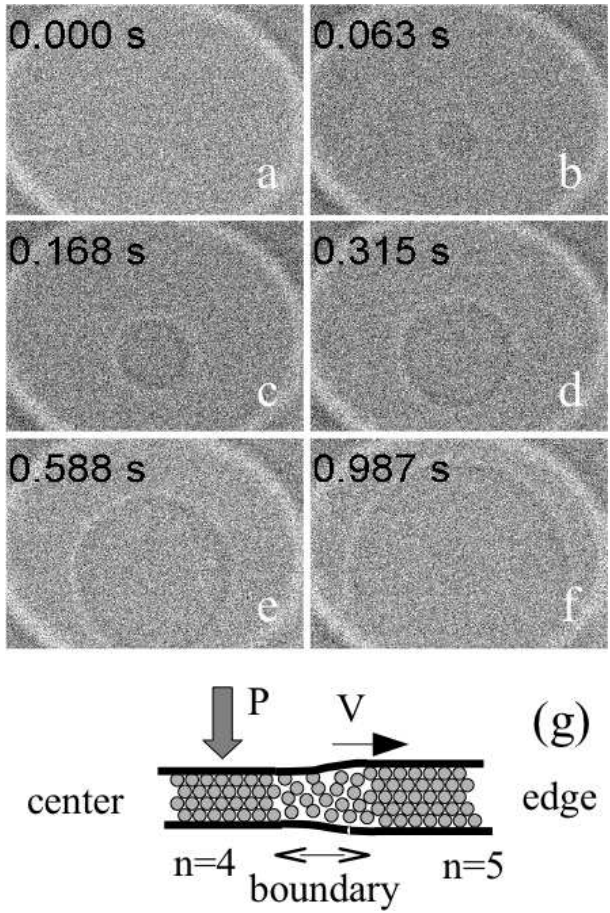}
$$
\caption{Pictures (height 96 $\mu$m and width 128 $\mu$m) of the contact zone during a transition from 5 to 4 layers of OMCTS. Time is stamped in the upper left corner of pictures $a$ to $f$. Driving velocity: 20 nm.s$^{-1}$. 
($g$): sketch of thickness variation of the film during squeeze-out.
}
\label{fig:fig3}
\end{figure}

When the applied pressure reaches some critical value, we note the apparition, close to the center of the contact, of a front which subsequently propagates across the contact (Fig. \ref{fig:fig3}). 
Although we cannot provide a detailed explanation for the origin of the observed contrast, we know from simultaneous thickness measurements using the fringes of equal chromatic order that
such a front delimits a zone of $n-1$ monolayers (behind the front) growing inside a film of thickness $n$ (ahead of the front).
The series of images presented on Fig. \ref{fig:fig3}, which illustrates a transition from 5 to 4 layers, therefore shows a ``squeeze-out front'', as observed by Becker and Mugele (BM) \cite{BM}. 

The apparent width of the front, which spans several microns, most probably results from multiple reflections in the locally curved Fabry-Perot cavity, and does not correspond to the physical width of
the disordered boundary zone which connects the regions of thickness $n$ and $n-1$ (see Fig. \ref{fig:fig3}g). This is supported by thickness measurements across the front, using the FECO, which indicate that the width of the boundary
is below the spatial resolution of the setup, namely $1.5\,\mu$m. Besides, one expects from continuum elasticity that bending of the mica sheets in the boundary zone occurs over a length on the order of the mica 
thickness ({\it i.e.} a few hundreds of nm), which is consistent with the fact that we cannot spatially resolve the front width.

\begin{figure}[htbp]
$$
\includegraphics[width=8cm]{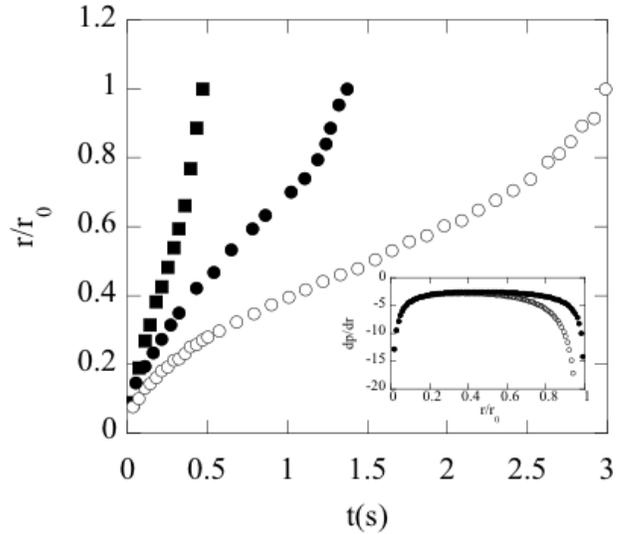}
$$
\caption{Time evolution of the the squeeze-out front position (plotted as the ratio of $r$ to $r_{0}$, the latter being the position of the contact edge). ($\blacksquare$): 
transtion from 7 to 6 layers. ($\bullet$): 
5$\rightarrow$4 transition. ($\circ$): 4$\rightarrow$3 transition. Inset : 2D pressure gradient as a fonction of $r/r_{0}$, computed from the PT model, using a Hertzian
 ($\bullet$) and a uniform ($\circ$) profile for the applied normal pressure.
}
\label{fig:fig4}
\end{figure}

We remark that, although propagation takes place in contacts which are frequently of elliptical shape \cite{footnote2}, the front remains circular during a large part of the 
growth phase. Loss of circularity is observed
only when the boundary line approaches close to the contact edge, where local acceleration induces the development of a bulge (Fig. \ref{fig:fig3}e). Such an acceleration is visible on Fig. \ref{fig:fig4}, where
we have plotted, as a function of time, the position $r$ of the front, for different transitions (7$\rightarrow$6, 5$\rightarrow$4, and 4$\rightarrow$3 layers). The position $r$ has been measured along the segment joining the nucleation site ($r=0$) to the point along the 
contact periphery which is first reached by the front ($r=r_{0}$). It can be seen that:

(i) front propagation is slower in thinner films,

(ii) for each transition, after an initial deceleration phase, the front reaches a constant velocity 
(respectively 70, 30 and 20 $\mu$m.s$^{-1}$ for 7$\rightarrow$6, 5$\rightarrow$4 and 4$\rightarrow$3 transitions), until $r/r_{0}\sim 0.6$ where it starts to accelerate. 

We now analyze in greater detail these results on squeeze-out front dynamics in order to (i) compare them quantitatively with those from BM \cite{BM}, and (ii) check their
 consistency with the results obtained from the drainage 
experiments described in section \ref{subsec:drain}. 

Following BM,  we interpret our results in the framework of a model proposed by Persson and Tosatti (PT), in which 
the transition of a confined film from $n$ to $n-1$ layers is described as follows \cite{PT}. When the applied pressure reaches a critical value, a patch of thickness $n-1$ 
layers nucleates in the film. Nucleation is accompanied by an elastic relaxation of the confining walls in the $n-1$ zone, which causes local bending of the 
surfaces (see Fig. \ref{fig:fig3}g). 
The elastic relaxation of the mica sheets is responsible for a 2D pressure gradient in the plane of the confined film, which drives liquid expulsion.
The PT model describes the growth of the squeezed area in the framework of 2D hydrodynamics, assuming that  resistance to propagation arises only from viscous drag due to flow ahead of the front:
\begin{equation}
\label{eq:NS}
\nabla p_{\text{2D}}=-\rho_{\text{2D}}\eta_{\text{eff}} v
\end{equation}
where $\nabla p_{\text{2D}}$ is the two-dimensional pressure gradient, $\rho_{\text{2D}}$ is the mass per unit area of the 2D ejected layer 
($\rho_{\text{2D}}=\rho a$, with $\rho$ the density and $a$ the
molecular diameter), $v$ is the front velocity, and $\eta_{\text{eff}}$ accounts for
viscous drag. Assuming fluid incompressibility and solving eq. (\ref{eq:NS}) in circular coordinates yields the following
equation for $A$, the area of the squeezed zone:
\begin{equation}
\label{eq:Adet}
\frac{dA}{dt}\ln\left(\frac{A}{A_{0}}\right)=-\frac{4\pi (p_{1}-p_{0})}{\eta_{\text{eff}}\rho_{\text{2D}}}
\end{equation}
where $A_{0}$ is the contact area, and $p_{1}-p_{0}$ is the 2D pressure drop across the front, which is related to the applied external pressure $P$: $p_{1}-p_{0}=aP$
 \cite{PT}. 

Integrating equation (\ref{eq:Adet}), PT express the time needed to completely expel one monolayer from the contact:
\begin{equation}
\label{eq:sqtime}
\tau= \frac{\rho_{\text{2D}}\eta_{\text{eff}}A_{0}}{4\pi (p_{1}-p_{0})}
\end{equation}

\begin{figure}[htbp]
$$
\includegraphics[width=8cm]{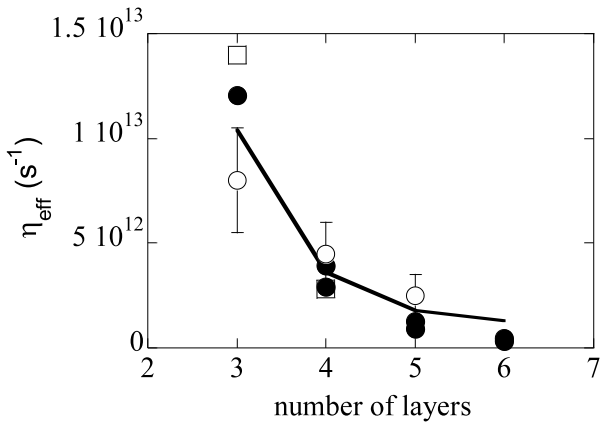}
$$
\caption{Effective drag coefficient $\eta_{\text{eff}}$ as a function of the number of OMCTS layers. ($\bullet$) $\eta_{\text{eff}}$ deduced from the measured squeeze-out time $\tau$.
($\circ$) results obtained by Becker and Mugele, taken from reference \cite{BM}.  (thick line) $\eta_{\text{eff}}$ calculated from  $\eta_{\text{ll}}= 0.2\times 10^{13}$s$^{-1}$ and 
$\eta_{\text{lw}}=8\times 10^{13}$s$^{-1}$. ($\square$) $\eta_{\text{eff}}$ deduced from inward propagation fronts (see section \ref{subsubsec:in}).
}
\label{fig:fig5}
\end{figure}

For a set  of $n\rightarrow n-1$ transitions taking place in a quasi-circular contact, we have measured the squeeze-out time $\tau$, and thus evaluated the effective viscosity coefficient $\eta_{\text{eff}}$
as a function of the number of layers $n$. Results are plotted on Fig. \ref{fig:fig5}, where we compare them with those obtained by BM. \\
We find a marked increase of $\eta_{\text{eff}}$ as the confined
film is made thinner. Moreover, it can be seen that our results are in good agreement with those of BM: both sets of results show the same trend and orders of magnitude for $\eta_{\text{eff}}$ as function of $n$.

Becker and Mugele further proposed that $\eta_{\text{eff}}$ should result from the combination of interlayer and layer/wall  viscous drag \cite{BM}. Assuming
that the 2D pressure is evenly distributed over the $n$ layers, and that equations similar to eq. (\ref{eq:NS}) hold for each layer $i$ in the film, they expressed $\eta_{\text{eff}}$ as a function of 
two drag coefficients $\eta_{\text{ll}}$ and $\eta_{\text{lw}}$ which control momentum transfer at the liquid/liquid and liquid/solid interfaces. \\
We have performed a similar analysis, and find that the best fit of our $\eta_{\text{eff}}(n)$ data is obtained with $\eta_{\text{lw}}=8\times 10^{13}$s$^{-1}$ and 
 $\eta_{\text{ll}}= 0.2\times 10^{13}$s$^{-1}$ (Fig. \ref{fig:fig5}). These values are in excellent agreement with those determined by BM, and account well for the observed 
dependence of $\eta_{\text{eff}}$ on $n$, which arises from the increasing contribution of $\eta_{\text{lw}}$ to the effective drag as the number of liquid/liquid 
interfaces decreases \cite{BM}. \\
Moreover, the value of $\eta_{\text{ll}}$ is very close to
$\eta_{\text{bulk}}/(\rho a^{2})\sim 0.25\times 10^{13}$, which is the drag coefficient obtained by assuming that sliding between adjacent liquid layers is controlled 
by bulklike viscous forces. 

We therefore come to the following conclusions:

(i) analysis of squeeze-out front dynamics agrees quantitatively with that of BM, and confirms their conclusion: the increase in effective viscosity of confined layered OMCTS 
results mainly from high friction between the walls and the two immediatly
adjacent liquid layers, while dissipation in the rest of the film remains bulk-like. 

(ii)  these results are  fully consistent with the analysis of flow curves presented in section \ref{subsec:drain}.

Let us now come back to the shape of the squeeze-out fronts. The PT model implicitly assumes that the line tension of the $n/(n-1)$ boundary does
not contribute to the front dynamics. Within the purely hydrodynamic description adopted in this model, it may then seem
surprising to observe that a squeeze-out front remains circular during a large part of its growth within an elliptical contact, whereas one would
expect a faster propagation along the minor axis of the contact, where pressure gradients are expectedly steeper. This feature of squeeze-out fronts was already visible in previous experiments by BM \cite{BM}, but no attempt was made to explain its origin. \\
Such an observation can actually be understood qualitatively within the framework of the PT model, by examining the radial dependence of the 2D pressure gradient which drives front propagation. 
From equations (\ref{eq:NS}) and (\ref{eq:Adet}), one obtains the following expression for the pressure gradient at $r=r_1$, $r_1$ being the front radius:
\begin{equation}
\label{eq:gradp}
\left.\frac{\partial p}{\partial r}\right\vert_{r=r_1}=\frac{aP}{r_{0}x\ln(x)}
\end{equation}
where $r_0$ is the contact radius, $P$ is the applied pressure, and $x=r_1/r_0$. 
On Fig. \ref{fig:fig4}, we have plotted the pressure gradient as a function $x$, assuming for $P$ a uniform pressure  ($P=P_0$ independent of $x$) or a Hertz pressure 
distribution inside the contact ($P=P_{\text{max}}\sqrt{1-x^2}$). In both cases, it can be seen that the 2D pressure gradient is nearly constant for $x$ ranging from 0.2 up to 0.6--0.7.
This agrees with our observation that the front dynamics is sensitive to the presence of the contact boundary only for $x\gtrsim 0.6$, while
the squeezed zone keeps its circular (nucleation) shape and grows at a quasi-constant velocity (see Fig. \ref{fig:fig4}) up to $x\lesssim 0.6$.

Now, it is noteworthy that, except for the late stage development of a bulge which can be attributed to the pressure gradient variation near the contact edge, squeeze-out fronts 
never exhibit roughening in our experiments. This suggests that the tension of the boundary line stabilizes circular front growth against the Saffman-Taylor fingering instability to be expected in
this geometry \cite{PM}. Now, line tension cannot be deduced from the analysis of front dynamics as presented above. However, we have observed a different scenario for squeeze-out, which 
we describe in the next section, from which we can make an estimate of the boundary line tension. This allows us to determine the length scale up to which line tension is 
important in describing front dynamics, and check the validity of the assumption made by Persson and Tosatti in their model.

\subsubsection{Inward fronts}
\label{subsubsec:in}

\begin{figure}[htbp]
$$
\includegraphics[width=8cm]{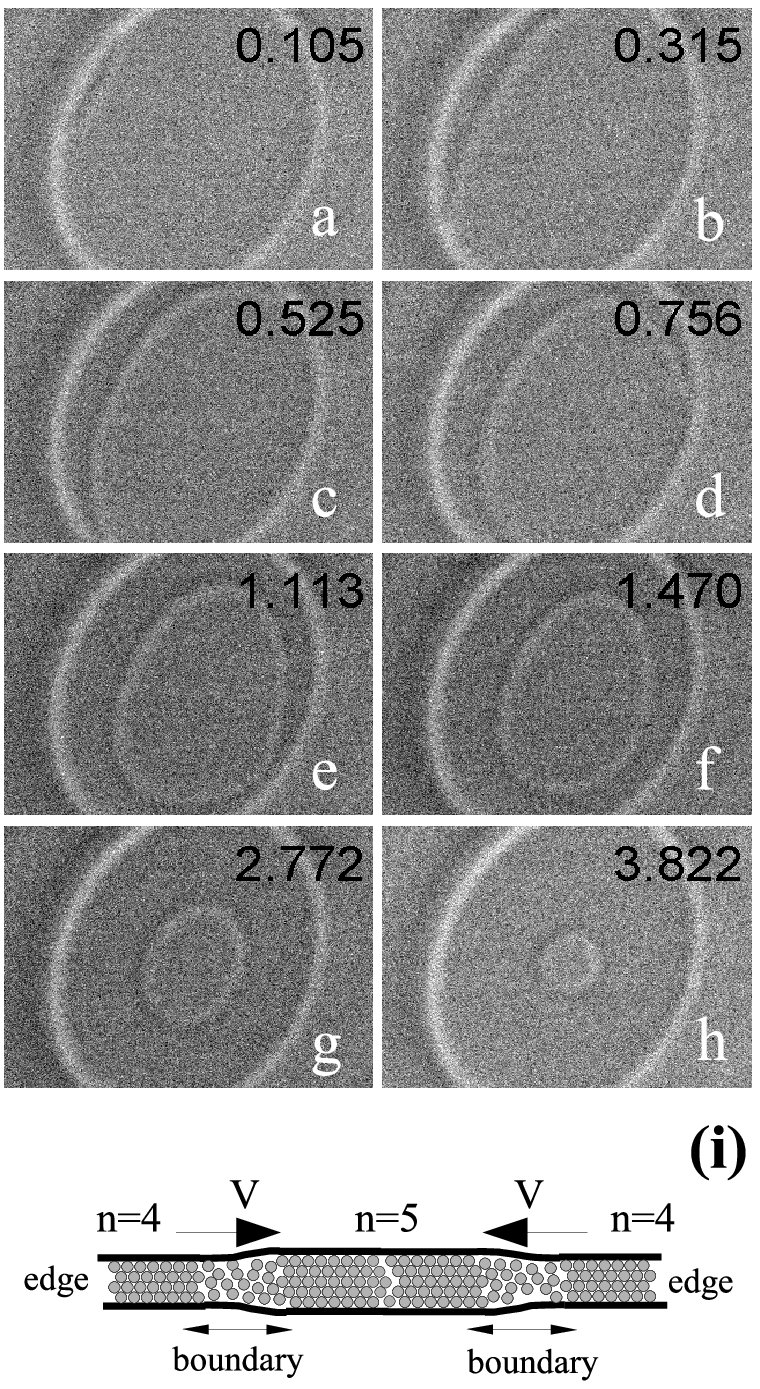}
$$
\caption{Pictures (height 96 $\mu$m and width 128 $\mu$m) of the contact zone during a transition from 5 to 4 layers of OMCTS, occuring via the inward propagation of a front. Time is stamped in the upper left corner of pictures $a$ to $h$. Driving velocity: 1 nm.s$^{-1}$. 
($i$): sketch of thickness variation of the film during squeeze-out.
}
\label{fig:fig6}
\end{figure}

In confinement experiments in which the driving speed is kept lower than, typically, $V=1$ nm.s$^{-1}$ ({\it i.e.} loading
rates $< 10^{-5}$~N.s$^{-1}$), we observe layering transitions which occur preferentially, as illustrated on Fig. \ref{fig:fig6}, by nucleation of a zone of $n-1$ layers
close to the contact edge, in contrast with what we have described in the previous section. Nucleation is followed by 
a rapid propagation {\it along} the contact periphery (Fig. \ref{fig:fig6}b-e), eventually leading to a closed loop which then propagates {\it towards the center} of the 
contact (see Fig. \ref{fig:fig6}e-h). Simultaneous thickness measurements show that the film is $n$ layer-thick {\it inside} the loop, and $n-1$ layer-thick outside. 
We therefore observe, when the confined film is slowly loaded, squeeze-out fronts which propagate {\it inward} (Fig. \ref{fig:fig6}i). \\
Moreover, when propagation takes place in an elliptical contact, we systematically observe that such inward fronts, which are also of elliptical shape during the major part of the squeeze-out process, become circular when they reach a characteristic 
radius $r_c\lesssim 5$--$10\,\mu$m (Fig. \ref{fig:fig6}h).

Let us first comment on the dependence of the nucleation/growth process on the loading rate. We attribute it to delayed elasticity of the glue layer employed to adhere the
 mica sheets to the cylindrical glass lenses. Indeed, the UV setting polyurethane glue we use in our experiments is a viscoelastic solid. Viscoelasticity manifests itself clearly when 
 separating the surfaces after an approach run during which the mica sheets were flattened: monitoring the shape of the interference fringes indicates that immediatly after 
 separation, the surfaces exhibit a radius of curvature which is larger than the initial one. The radius of curvature then relaxes back to its initial value, within a few tens of minutes.\\
Although the viscoelastic properties of the glue have not been characterized in detail, the following qualitative picture is likely to explain our observations. 
Nucleation of a $n-1$ layer-thick zone into a $n$ layer-thick film is expected to occur when the external pressure reaches some critical value inside the contact \cite{PT}. 
The time-dependent response of the glue layer most probably affect the pressure distribution in the contact during loading \cite{KLJ}. Under large enough loading rates, 
the response of the glue layer is mainly controlled by its instantaneous elastic modulus, which results in a quasi-Hertzian normal stress profile, {\it i.e.} displaying a  
pressure maximum close to the center of the contact. Nucleation
is therefore favored at the location of this maximum, and is accompanied by 2D bending of the mica sheets. Now, at low loading rates,  a long time is required before 
reaching the critical nucleation pressure. During this loading phase (where several tens of minutes may elapse between two transitions), viscoelasticity of the glue most likely 
results in a flattening of the pressure profile \cite{KLJ,LR}. The applied normal stress therefore becomes more uniformly distributed over the contact and does not display a 
clear maximum, in contrast to the Hertz-like situation. Under such conditions, nucleation preferentially occurs where bending of the mica sheets is most easy. Such is the case at the 
periphery of the contact, where quasi-1D bending along the contact boundary is easier than 2D bending at the center.  

Let us now focus on the late stage of inward front propagation. It is seen on Fig. \ref{fig:fig6}h that an elliptical front becomes circular when its size is smaller 
than $r_c\simeq$10 $\mu$m. This indicates that the line tension of the $n/(n-1)$ boundary becomes relevant in its dynamics on length scales below $r_{c}$. The assumption made 
by Persson and Tosatti in their model is therefore valid for fronts of radius $r>r_{c}$, which is the case during the major part of front propagation in our experiments. This gives
support to the analysis of squeeze-out dynamics within the framework of the PT model performed in section \ref{subsubsec:out}.

As already discussed by Persson and Mugele \cite{PM}, the line tension $\Gamma$ contains two contributions: one ($\Gamma_{\text{int}}$) is associated with 
disorder in the $n/(n-1)$ boundary zone, where the film thickness is not an integer number of monolayers, and the other ($\Gamma_{\text{el}}$) comes from 
the elastic line energy due to bending of the mica sheets. Assuming, as we have already discussed earlier in section \ref{subsubsec:out}, that distortion of the mica extends 
laterally over a length scale on the order of the thickness of the sheets, the elastic line tension is given by: $\Gamma_{el}\simeq Ea^2/\left[2 \pi(1-\nu^{2})\right]$ 
\cite{PM}, where $a$ is the molecular diameter, $E$ and $\nu$ are respectively the Young Modulus and the Poisson ratio of the sheets.
Taking $E\simeq 18$ GPa, $\nu\simeq 0.45$ \cite{PT}, and $a\simeq 10^{-9}$~m, one thus estimates $\Gamma_{\text{el}}\simeq 3.10^{-9}$~N.\\
From our observations, we can perform a direct evaluation of the line tension. Indeed, the circularization of an elliptical inward front
indicates that the 2D pressure drop which drives the flow, namely $p_1-p_0=aP$, becomes comparable with the 2D Laplace pressure:
\begin{equation}
\label{eq:recirc}
aP\sim \frac{\Gamma}{r_c}
\end{equation}
With $a\simeq 10^{-9}$~m, $r_c\simeq 10\, \mu$m and $P\simeq 400$~kPa (the values of the critical radius and the average pressure for the transition illustrated on 
Fig. \ref{fig:fig6}), we get $\Gamma \simeq 4.10^{-9}$~N. This value is in strikingly good agreement with that of $\Gamma_{\text{el}}$ given above, and leads to conclude 
that $\Gamma_{\text{el}}\gg \Gamma_{\text{int}}$. This constitutes, 
to our best knowledge, the first direct evidence that the line tension of a squeeze-out front arises mainly from bending of the confining surfaces, and confirms what was guessed 
from previous numerical studies \cite{PM}.

The above analysis further calls for a remark regarding the order of magnitude of the critical nucleation radius. In their article, Persson and Tosatti 
 derive an expression for the nucleation energy which arises from three contributions: a line energy ($2\pi r \Gamma_{\text{int}}$, with $\Gamma_{\text{int}}$ the tension 
 resulting from disorder in the boundary zone only), a change in internal energy of the film, and a third term associated with elastic relaxation of the confining walls. 
 Minimizing this nucleation energy with respect to the radius of the squeezed patch, they deduce a critical radius $r_c$ of approximately 1 nm. 
 However, following the above analysis concerning the origin of line tension, we express the condition
 for nucleation as:
\begin{equation}
\label{eq:nucl}
\pi r_{c}^{2}Pa = 2\pi r_c (\Gamma_{\text{el}}+\Gamma_{\text{int}})\simeq 2\pi r_{c} \Gamma_{\text{el}}
\end{equation}
Such a criterion is analogous to the nucleation condition of an elementary dislocation loop in self-supported smectic films, which exhibit thinning transitions qualitatively 
similar to those of confined OMCTS films \cite{osw1,osw2}. \\
Taking for $\Gamma_{\text{el}}$ the value determined above, we obtain a critical nucleation radius on the order of a few {\it micrometers}, considerably larger 
than the estimate made by Persson and Tosatti. We believe that such a difference arises from the fact that PT calculate the contribution from elastic relaxation  
for infinitely thick confining walls, and thus neglect the plate-bending contribution responsible for $\Gamma_{\text{el}}$ \cite{PT}. 

Let us finally come back to the inward front dynamics. As in the case of outward fronts, the bidimensional pressure gradient induced by bending of the mica sheets drives the 
squeeze-out process. However, as opposed to the situation described in section \ref{subsubsec:out}, inward fronts imply that the thicker (n layers) part of the film has to flow
into the surrounding thinner (n-1 layers) part. Two mechanisms should in principle control the squeeze-out dynamics in such a case: (i) shear flow in the (n-1) layer-thick 
part of the film, and (ii) permeation flow, normal to the plane of the layers, allowing for interlayer transport of molecules in the thick zone ahead of the front. These two 
mechanisms were proposed to explain the spreading dynamics of stratified droplets \cite{PGGcaza}. We evaluate the role played by the permeation process as follows.
Under the assumption that only shear flow is relevant in inward fronts dynamics, we estimate, as in section \ref{subsubsec:out}, the effective drag coefficient,
$\eta_{\text{eff}}$, from the squeeze-out time $\tau$ for different thinning transitions. As shown on Fig. \ref{fig:fig5}, we find that $\eta_{\text{eff}}$ for 
an ``inward'' $n\rightarrow n-1$ transition agrees well with $\eta_{\text{eff}}$ measured for an ``outward'' $n-1\rightarrow n-2$ transition.
In both situations, flow of molecules between the front and the contact periphery occurs through a channel of thickness $n-1$. \\
Such an agreement between $\eta_{eff}$ for outward and inward fronts leads us to conclude that inward propagation is mainly governed by shear 
flow between the front and the contact edge, and that, in our confined configuration, the permeation mechanism mentioned above contributes negligibly to dissipation.

\subsection{Rapidly quenched films: layer-by-layer spinodal thinning}
\label{subsec:spin}

Eventually, we present results from experiments in which  OMCTS is confined using driving velocities in the range 0.5--10 $\mu$m.s$^{-1}$, {\it i.e.} at least three orders of magnitude larger than those employed in sections \ref{subsec:drain} and \ref{subsec:fronts}. 
Mica sheets of thickness on the order of a few hundreds of nanometers are also employed here, in order to use the imaging method described in section \ref{subsec:fronts}, 
and observe whether, under such loading conditions, thinning still occurs via discrete expulsion of single molecular layers. 

The mica surfaces are approached until the normal force reaches approximately 15--20 mN, which yields a final applied pressure $P\gtrsim 1$ MPa. 
Under such a pressure, quasi-statically loaded OMCTS films are 3-layer-thick. In the present experiments,
the normal stress applied on the confined film increases from 
zero to the final pressure within a time ranging from 0.25 to 5 seconds (as opposed to tens of minutes under slow confinement conditions), depending on the driving velocity. 
In this velocity range, our observations do not appear to be sensitive to the speed of the pressure quench. Moreover, we have performed experiments starting from both 
relatively thick ($\sim 100$ nm) and
thin layered films ($\sim$6--7 monolayers), and did not notice any influence of the initial film thickness on the behavior we now describe.

\begin{figure}[htbp]
$$
\includegraphics[width=7cm]{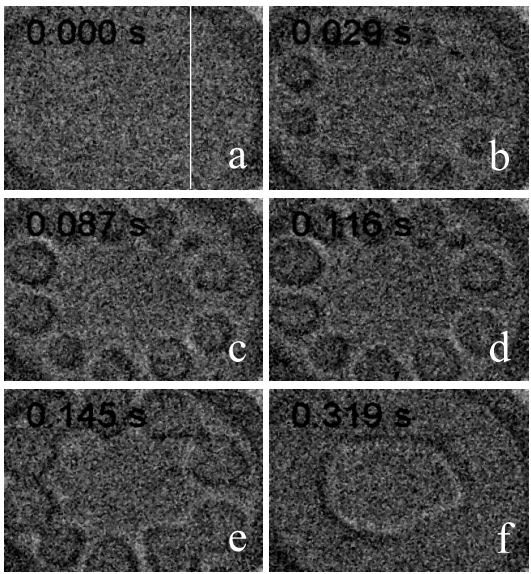}
$$
\caption{Pictures (96 $\mu$m$\times$128 $\mu$m) of the contact zone during a transition from 5 to 4 layers, occuring via spinodal decomposition. 
Time is stamped in the upper left corner of pictures $a$ to $f$. The vertical white line on (a) indicates the line along which thickness measurements are performed using the FECO.}
\label{fig:fig7}
\end{figure}

\begin{figure}[htbp]
$$
\includegraphics[width=7cm]{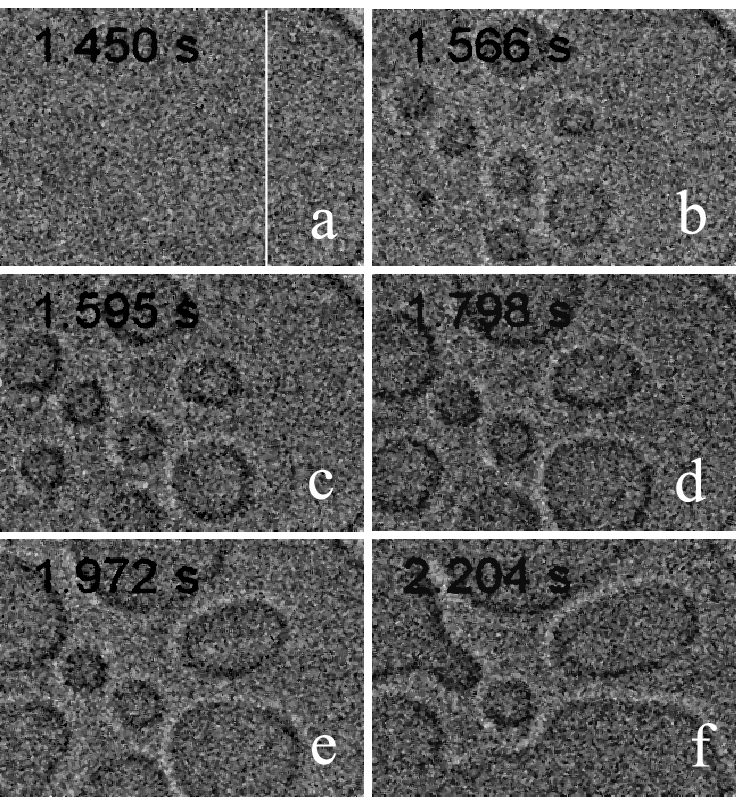}
$$
\caption{Pictures (96 $\mu$m$\times$128 $\mu$m) of the contact zone during a transition from 4 to 3 layers, occuring via spinodal decomposition. Time is stamped in the 
upper left corner of pictures $a$ to $f$.  The vertical white line on (a) indicates the line along which thickness measurements are performed using the FECO.
}
\label{fig:fig8}
\end{figure}

\begin{figure}[htbp]
$$
\includegraphics[width=8cm]{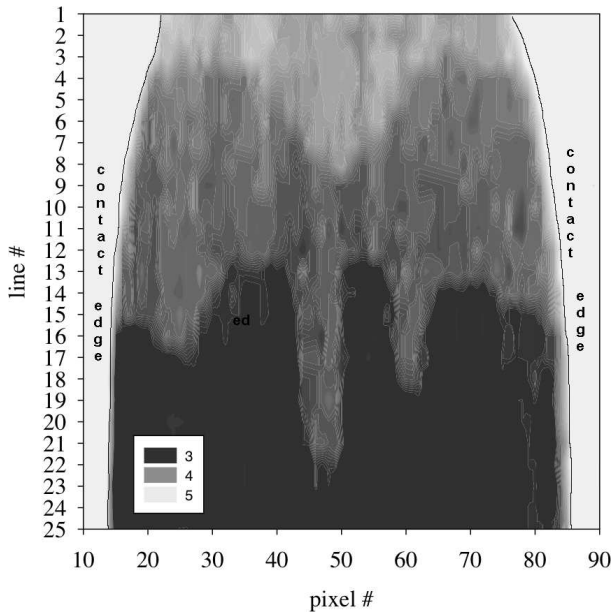}
$$
\caption{Spatiotemporal evolution of the film thickness measured along the the line drawn on Fig. \ref{fig:fig7}a and
\ref{fig:fig8}a. The film thickness (in number of OMCTS layers) is coded in gray level, as shown in the figure inset.  Lines (vertical axis) are separated by 150 ms (time
 increases from line 1 to line 25), and pixels
(horizontal axis) represent 2 $\mu$m. Line 1 corresponds  to image (a)  on Fig. \ref{fig:fig7}. 
Line 11 corresponds  to image (a)  on Fig. \ref{fig:fig8}. The late stage of the $5\rightarrow 4$ transition is seen between lines 3 and  8, which corresponds, after 
merging of the 4-layer-thick zones, to the receeding of a 5-layer-thick central part. Measurements during the $4\rightarrow 3$ transition (lines 11 to 22) 
more clearly show the coexistence of multiple thin regions growing  into the contact zone.
}
\label{fig:fig9}
\end{figure}

When the applied pressure is rapidly increased up to $\sim 1$ MPa, we find that films thin down to three monolayers, {\it i.e.} down to a thickness equal to
 that reached under slow loading.
However, in contrast to the situation of slow normal loading, we observe that for thicknesses on the order of 6--7 nm and below, drainage takes place by the following succession of 
sequences. Within the contact zone where mica surfaces are flattened, multiple circular patches, a few microns in diameter, 
appear quasi-simultaneously (Fig. \ref{fig:fig7}b and \ref{fig:fig8}b). These patches subsequently grow (Fig. \ref{fig:fig7}b-d and \ref{fig:fig8}b-e), until they 
merge (Fig. \ref{fig:fig7}e and \ref{fig:fig8}f) and eventually leave a single tortuous front which then propagates through the contact. We have observed the repetition of 
such a sequence up to four times during a single squeeze-out run (the last two of these are illustrated on Fig. \ref{fig:fig7} and \ref{fig:fig8}). We also note that, from one sequence to 
the next one, circular patches appear at different locations inside the contact (see Fig. \ref{fig:fig7}b and \ref{fig:fig8}b).\\
Furthermore, thickness measurements along a line across the contact show that each of these sequences corresponds to drainage of OMCTS by expulsion of {\it one single 
monolayer}: starting from a film of uniform thickness $n$ layers ($n=5$ on Fig \ref{fig:fig7}a and $n=4$ on Fig. \ref{fig:fig8}a), patches of thickness $n-1$ 
appear and grow until one monolayer is completely expelled from the contact. Then, starting from the resulting film of uniform thickness $n-1$, the same scenario 
repeats itself, down to $n=3$. Such a time evolution of the film thickness, measured during the sequences illustrated on Fig \ref{fig:fig7} and Fig. \ref{fig:fig8}, is shown 
on the spatio-temporal plot of Fig. \ref{fig:fig9}.

The first conclusion to be drawn from these experiments is that even under conditions of extremely rapid loading, OMCTS molecules keep their tendency to form layers parallel 
to the solid walls, which results, under confinement, into  
stepwise drainage of the fluid.

Now, the mechanism for the layer-by-layer thinning transitions is obviously affected by the loading conditions. Patterns rather similar to those described in this section were 
occasionally observed in previous studies (see for instance Fig. 31 in ref. \cite{PM}), but were not discussed in detail.
In analogy with first-order phase transitions, the dynamics of which is controlled by the level of supersaturation \cite{cahn,binder}, we propose the following qualitative picture to rationalize our observations. \\
The free energy of a confined simple fluid is an oscillating function of its thickness, with a 
periodicity comparable to the size of a molecule \cite{GLL}. Such an energy landscape is biased by the application of an external pressure.
When the pressure applied to a film of $n$ molecular layers is {\it slowly} increased, the energy barrier separating metastable states of thickness $n$ and $n-1$ layers is progressively 
lowered. This barrier is eventually overcome under the combined effect of pressure and thermal activation. This results in nucleation of a zone of thickness $n-1$ layers into the 
film, which subsequently grows, as described in section \ref{subsec:fronts}.\\
Conversely, when the normal force is {\it rapidly} increased, films of thickness $n>3$ layers are submitted to an instantaneous pressure
larger than the critical pressure which would be required for thinning via nucleation. This leads to a pressure-induced bias of the solvation energy  large enough to erase
nucleation barriers, {\it i.e} this leads to highly ``supersaturated'' films.
Films of thicknesses larger than 3 monolayers are therefore 
{\it unstable}. The apparition, followed by growth and coalescence of multiple thin zones in such unstable films further suggests that thinning transitions occur by 
amplification of thickness fluctuations which subsequently grow inside the contact zone. 

We therefore conclude, in analogy with spinodal dewetting of ultrathin supported liquid films \cite{dewet}, that confined films submitted to the sudden application of a large 
pressure undergo thinning transitions by a 
mechanism akin to spinodal decomposition. 

One may of course wonder why spinodal thinning occurs via such a layer-by-layer process. This is most likely attributable to the
 elastic cost due to local bending of the mica sheets at the boundary of the thinner zones, as discussed in section \ref{subsubsec:in}. Indeed, the line tension is expected 
 to increase with increasing height difference, 
 which probably hinders the development of thickness variations of more than one monolayer.

\begin{figure}[htbp]
$$
\includegraphics[width=7cm]{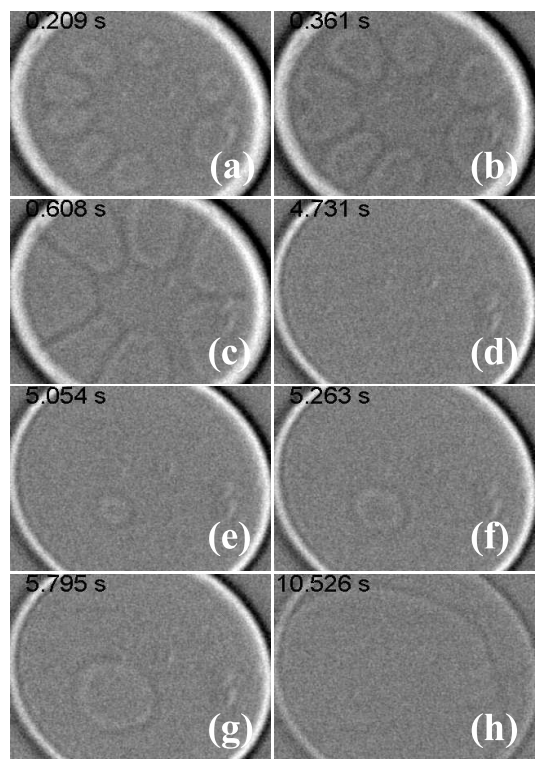}
$$
\caption{Pictures (96 $\mu$m$\times$128 $\mu$m) of the contact zone during a transition from 5 to 4 layers, occuring via spinodal decomposition ($a$ to $d$), followed by a transition from 4 to 3 layers taking place by 
nucleation/growth ($d$ to $h$). Time is stamped in the upper left corner of the pictures.
}
\label{fig:fig10}
\end{figure}

Our interpretation is further supported by the following experiment. Starting from a film of initial thickness $n=6$ layers, we increase the pressure, 
within 0.25 s, up to $P\simeq 2$ MPa, {\it i.e.} twice the final pressure reached in the experiment illustrated on Fig. \ref{fig:fig7} and \ref{fig:fig8}. Under such loading 
conditions, we observe that the first 
3 transitions ($6\rightarrow 5$, 
$5\rightarrow 4$ and $4\rightarrow 3$)  still occur by spinodal-like decomposition, but we observe one more transition ($3\rightarrow 2$) taking place by nucleation of a single 
thin zone which then propagates outward. 
A sequence of images of the contact area during the last two transitions ($4\rightarrow 3\rightarrow 2$) is given on Fig. \ref{fig:fig10}. \\
These observations are consistent with the fact that supersaturation controls the thinning dynamics. Indeed,
 the nucleation pressure is higher in thinner films, and thus the difference between the instantaneous applied pressure and the
nucleation pressure decreases as a film thins down. Under a given applied pressrue, the level of supersaturation therefore
decreases as the film thickness decreases. Fig. \ref{fig:fig10}  indicates that, under $P\simeq$2 MPa, the pressure-induced bias is such that nucleation barriers for transitions 
down to $4\rightarrow 3$ are erased, whereas the energy barrier for the $3\rightarrow 2$ transition is sufficiently lowered for nucleation to be observed. The thinning dynamics 
thus crosses over from spinodal
 decomposition to nucleation/growth as supersaturation decreases, as expected in first order transition kinetics.

\section{Conclusions}
\label{sec:conclusion}

We have studied in detail the effect of confinement rate on the squeeze-out dynamics of the simple fluid OMCTS. For this purpose, we have performed, in parallel, measurements of approach curves and imaging of the confined liquid films.

 Analysis of the force-distance curves during approach at various rates have shown that:\\
(i) confined films of OMCTS exhibit layering even under rapid confinement, \\
(ii) if one accounts for two quasi-immobile monolayers in contact with the confining walls, viscous drag in layered films is essentially bulklike, {\it i.e.} confined OMCTS 
retains its bulk viscosity.\\
Such a behaviour is in strong contrast with that of a linear alkane, n-hexadecane, which, when rapidly quenched, does not exhibit layering and displays an enhanced
 viscosity \cite{LB1}. This points to the importance of the molecular architecture of a confined lubricant on its flow properties under confinement. 

Besides, imaging of the confined films, using a very simple technique, has allowed us to show that:  \\
(i) under low or moderate loading rates, layering transitions (thinning from $n$ to $n-1$ monolayers) occur by nucleation and growth of a thin patch within the film. \\
(ii) when the  applied pressure is increased abruptly, the film is unstable and undergoes a series of layer-by-layer spinodal decompositions.\\
We thus provide evidence that the stability and thinning mechanism of confined films of OMCTS  are controlled by the loading rate and the magnitude of the applied pressure,
 in qualitative analogy with phase separation in pressure quenched solutions \cite{solpol} or dewetting of thin liquid films \cite{dewet}. 

Furthermore, we have found that the above mentioned nucleation/growth mechanism may happen following two scenarii. These results suggest that nucleation of a thin zone in a confined film is controlled by the local pressure on the film and by bending of the confining sheets:\\
(i) when the pressure applied on the film is close to uniform, nucleation occurs where bending of the confining walls is favored, {\it i.e.} along the perimeter of the contact. 
Thinning  of the film further proceeds by propagation of a front towards the center of the film. This type of propagation has allowed us to make the first quantitative estimate of 
the line tension of the boundary separating zones of different thicknesses in a confined film. \\
(ii) when the profile of the applied pressure exhibits a well defined maximum, nucleation occurs at the location of this maximum, and thinning proceeds by propagation of a 
squeeze-out front from the center towards the edge of the confined film. These observations are in excellent agreement with those of Mugele {\it et al} \cite{MS,BM}, who 
first reported on such squeeze-out fronts. Moreover, analysis of front dynamics, within the framework of a model initially proposed by Persson and Tosatti, leads us to a 
conclusion similar to that of Mugele. The viscous drag which opposes front propagation
results from two contributions: high friction at the interfaces between the walls and the immediately adjacent liquid layers, and bulklike friction in the rest of the film. 

This is fully consistent with what we conclude from our analysis of the flow curves.
To our best knowledge, this is the first report of such a consistency check between different types of measurement using the Surface Forces Apparatus.     

\acknowledgments{We are grateful to Tristan Baumberger and Christiane Caroli for fruitful and stimulating discussions.}



\begin{thebibliography}{srt}
\bibitem{footnote} We do not notice any influence of the exact thickness of the mica sheets on the phenomenology we describe in sections \ref{subsec:fronts} and \ref{subsec:spin}. Actually, one single thin mica sheet (the other being of micrometric thickness) is enough to obtain the results reported here.
\bibitem{footnote2} Such an elliptical shape of the contact zone is due to the fact that the cylindrical lenses supporting the mica sheets are not perfectly crossed.
\end{thebibliography}
\end{document}